\documentclass{acm_proc_article-sp}
\pdfoutput=1

\usepackage{times}
\usepackage{xspace,amssymb,mathtools,fixltx2e}
\usepackage[T1]{fontenc}
\usepackage{listings}
\usepackage{amsmath, bbm}
\usepackage{pgf}
\usepackage[latin1]{inputenc}
\usepackage{booktabs} 
\usepackage{verbatim}
\usepackage{semantic}
\usepackage{multirow}
\mathlig{-><-}{\rightarrow\leftarrow}
\usepackage{stmaryrd}
\usepackage{float}
\restylefloat{table}
\usepackage[breaklinks=true, bookmarksopen=true,bookmarksnumbered=true]{hyperref}

\newcommand{\footnoteUrl}[1]{\footnote{\url{#1}}}

\definecolor{tealJ}{rgb}{0.5,0.8,0.7}
\definecolor{tealX}{rgb}{0.5,0.7,0.8}

\lstdefinestyle{Sparql}{numberblanklines=true, 
    morekeywords={SELECT, FROM, WHERE, FILTER, GROUP BY, IN, AS, LIMIT,
    OFFSET,PREFIX,OPTIONAL,UNION,foaf,rdf,dct,skos,rdfs,ex,xsd,owl,skosxl,doap,void,dbo,org,cex, qp}}

\lstdefinestyle{Turtle}{
    numberblanklines=true, 
    morekeywords={prefix, @prefix 
    , foaf, rdf, skos, skosxl, rdfs, ex
    , xsd, wr, wt, dc, lemon, doap, doap
    , void, dbo, owl, a, wf, qb, dct, org, cex
    }}

\lstdefinestyle{ShExC}{
  morecomment=[s][\color{tealX}]{\%GenX\{}{\%\}},
  morecomment=[s][\color{tealJ}]{\%GenJ\{}{\%\}},
  morecomment=[s][\color{violet}]{\%js\{}{\%\}},
  morecomment=[s][\color{orange}]{\%sparql\{}{\%\}},
  keywords={start,VIRTUAL,CLOSED,EXTRA,IRI,BLANKNODE,PATTERN,
  , prefix
  , :Country, :DataSet, :Slice, :Observation, :Computation, :Indicator, :Organization
  , foaf, rdf, skos, skosxl, rdfs, ex, a, sh
  , xsd, wr, wt, dc, lemon, doap, doap
  , void, dbo, owl, wf, qb, org, dct, cex, qp},
  ndkeywords={class, export, boolean, throw, implements, import, this},
  ndkeywordstyle=\color{darkgray}\bfseries,
  identifierstyle=\color{black},
  sensitive=false,
  commentstyle=\color{purple}\ttfamily,
  stringstyle=\color{red}\ttfamily,
  morestring=[b]',
  morestring=[b]"
}

\lstdefinestyle{Sparql}{numberblanklines=true, 
    morekeywords={SELECT, FROM, WHERE, CONSTRUCT, FILTER, GROUP BY, IN, AS, LIMIT,
    OFFSET,PREFIX,OPTIONAL,UNION
    }}

\lstdefinestyle{SQL}{numberblanklines=true, 
    morekeywords={CREATE,TABLE,ENUM,FOREIGN,KEY,ENUM,REFERENCES}}

\lstdefinestyle{XML}
{
  morestring=[b]",
  morestring=[s]{>}{<},
  morecomment=[s]{<?}{?>},
  stringstyle=\color{black},
  identifierstyle=\color{darkblue},
  keywordstyle=\color{cyan},
  morekeywords={xmlns,version,type}
}

\lstdefinelanguage{scala}{
  morekeywords={abstract,case,catch,class,def,%
    do,else,extends,false,final,finally,%
    for,if,implicit,import,match,mixin,%
    new,null,object,override,package,%
    private,protected,requires,return,sealed,%
    super,this,throw,trait,true,try,%
    type,val,var,while,with,yield},
  otherkeywords={=>,<-,<\%,<:,>:,\#,@},
  sensitive=true,
  morecomment=[l]{//},
  morecomment=[n]{/*}{*/},
  morestring=[b]",
  morestring=[b]',
  morestring=[b]"""
}

\begin{document}

\title{Validating and Describing Linked Data Portals using Shapes}
\numberofauthors{2} 
\author{
\alignauthor
Jose Emilio Labra Gayo\\
       \affaddr{University of Oviedo}\\
       \affaddr{Dept. of Computer Science}\\
       \affaddr{C/Calvo Sotelo, S/N}\\
       \email{labra@uniovi.es}
\alignauthor
Eric Prud'hommeaux\\
\affaddr{World Wide Web Consortium (W3C)}
\affaddr{MIT, Cambridge, MA, USA}
\email{eric@w3.org} \\
\and
\alignauthor 
Harold Solbrig\\
       \affaddr{Mayo Clinic}\\
       \affaddr{College of Medicine, Rochester, MN, USA}\\
\alignauthor
Iovka Boneva\\
       \affaddr{University of Lille}\\
}

\maketitle

\begin{abstract}
Linked data portals need to be able to advertise and describe the structure of their content.
A sufficiently expressive and intuitive ``schema'' language will allow portals to communicate these structures.
Validation tools will aid in the publication and maintenance of linked data and increase their quality.

Two schema language proposals have recently emerged for describing the structures of RDF graphs: 
 Shape Expressions (ShEx) and Shapes Constraint Language (SHACL).
In this paper we describe how these formalisms can be used in the development 
of a linked data portal to describe and validate its contents. 
As a use case, we specify a data model inspired by the WebIndex data model, a medium size linked data portal, 
using both ShEx and SHACL, and we propose a benchmark that can generate compliant test data structures of any size. 
We then perform some preliminary experiments showing performance of one validation engine based on ShEx. 
\end{abstract}


\section{Introduction}
\label{sec:Introduction}

Linked data portals have emerged as a way to publish data on the Web following a set of principles~\cite{bernerslee06} which 
improve data reuse and integration. 
As indicated in~\cite{bizer_linked_2009}, linked data relies on documents using RDF representations to make statements that link arbitrary things in the world.
RDF serves as a data integration language and, for some linked data applications, a database technology or interoperability layer.
However, there is a lack of an accepted practice for declaring the structure (shape) of an RDF graph in a way that can be automatically validated before and after publishing it~\cite{SHACLCharter}. 
Linked data projects involve several stakeholders at different stages and with different roles. 
When developing a linked data portal, domain experts and web developers need some way to communicate the data model and the RDF 
representations that they will produce. 
The potential consumers of linked data also need to easily understand the structure of the RDF data and tools to reliably validate the 
data with respect to that structure before consuming it. 
We consider that the overall quality of linked data portals improves when there is an intuitive tool that can be used to declaratively specify and communicate the data model.

Validation is standard practice in conventional data languages.
Today's software development methodologies use a variety of models 
including DDL constraints for SQL databases and 
XML Schema or RelaxNG for XML documents. 

In the last few years, there has been an increased interest in creating technologies that enable
 the description and validation of RDF data structures. 
In 2013, an RDF validation workshop was organized by the W3C to gather the requirements of the different stakeholders. 
 A conclusion of the workshop was that, although SPARQL could be used to validate RDF, there was a need for a terse, higher level language. 
Shape Expressions (ShEx) emerged as such a language, intended to perform the same function for RDF graphs as Schema languages do for XML~\cite{EricSemantics2014}. 
ShEx was designed as a high-level, concise language intended to be human-readable using a Turtle-like syntax familiar to users of regular expression based languages like RelaxNG or XML Schema. 

In 2014 the W3C chartered the RDF Data Shapes working group to produce a language for defining structural constraints on RDF graphs~\cite{SHACLCharter}. 
The language, called SHACL (SHApes Constraint Language), serves a similar purpose as Shape Expressions. 
In this paper we describe how those two languages can be used to describe the contents of linked data portals in a way that their content can be automatically validated. 
At the time of this writing, both ShEx and SHACL are still work-in-progress and their implementations are mainly 
proof-of-concept.
Our description of SHACL is based on the Working Draft published on January 2016\footnote{\url{http://www.w3.org/TR/2016/WD-shacl-20160128/}}.

We use the WebIndex data portal\footnote{\url{http://thewebindex.org/}} as a test case.  WebIndex is a linked data portal of medium size (around 3.5 million of triples) that contains an interesting data model of statistical information with interrelated shapes and reuses several existing vocabularies like 
RDF Data Cube, Organization Ontology, Dublin Core~, etc.
It was selected because it is based on a real linked data portal that was designed by one of the authors of this paper. 
The documentation of the original data model was in fact described using an early version of ShEx~\footnote{\url{http://weso.github.io/wiDoc}}.
In this paper we use a variation of the original data model in order to better describe some modelling features 
and to avoid some repetitive features. At the end of section~\ref{sec:ShapeExpressions} we enumerate the main differences between both data models.

\paragraph{Previous Work and Contributions}
A first version of ShEx was presented at the Semantics conference~\cite{EricSemantics2014}.
In~\cite{Slawek2015} we studied the complexity of Shape Expressions for open and closed shapes without negation. 
We  showed  that  in  general  the  complexity  is NP-complete, identified tractable fragments, 
and proposed a validation algorithm for a restriction called deterministic single-occurrence shape expressions. The semantics was defined in terms of regular bag expressions. In~\cite{LabraShexDerivatives15} we described a validation algorithm 
 employing shape expression derivatives that was more efficient than backtracking and in a recent paper~\cite{BonevaShapeExpressionSchemas15} we have presented a well founded semantics of ShEx with negation and recursion as well as a full validation algorithm and some guidelines for efficient implementation.

This paper is based on a previous paper that we presented at the 1st Workshop on Linked Data Quality~\cite{LabraLDQSemantics}. 
In that paper we presented only a description based on ShEx while in this paper we present both ShEx and SHACL. 
In this paper we also present a tool that generates WebIndex data with random values 
of any size that can be used as a benchmark.  
We have also added a first performance assessment of one of the ShEx implementations using that tool.

\paragraph{Structure of the paper} The paper is organized as follows. Section~\ref{WebIndexDataModel} describes the WebIndex data model in an informal way. 
Section~\ref{sec:ShapeExpressions} describes the data model using ShEx in a formal way. 
It also describes some ShEx tools and approaches to validate linked data portals. 
Section~\ref{sec:ValidatingWithSHACL} describes the same data model using SHACL and also describes some
 SHACL implementations.
Section~\ref{sec:ComparingShExSHACL} presents some features that were not used for the WebIndex data model 
 and gives an overview of some of the differences between ShEx and SHACL.
Section~\ref{sec:PerfomanceComparison} describes wiGen, a tool to generate WebIndex-like data which 
 can be used to develop some performance tests. We describe some experiments that we have done to 
 evaluate one ShEx engine. Finally, section~\ref{sec:Related} presents related work and we present some conclusions in section~\ref{sec:Conclusions}.
\section{WebIndex data model}
\label{WebIndexDataModel}

WebIndex is a multi-dimensional measure of the World Wide Web's contribution to development and human rights globally. It covers 81 countries and incorporates indicators that assess several areas like universal access; freedom and openness; relevant content; and empowerment. 

The first version of WebIndex offered a data portal where the data was obtained by transforming raw observations and precomputed values 
from Excel sheets to RDF. 
The second version employed a validation and computation approach that published a verifiable version of the Web Index data~\cite{LabraMTSR14}.

\begin{figure*}[h]
\begin{center}
\includegraphics[width=\textwidth]{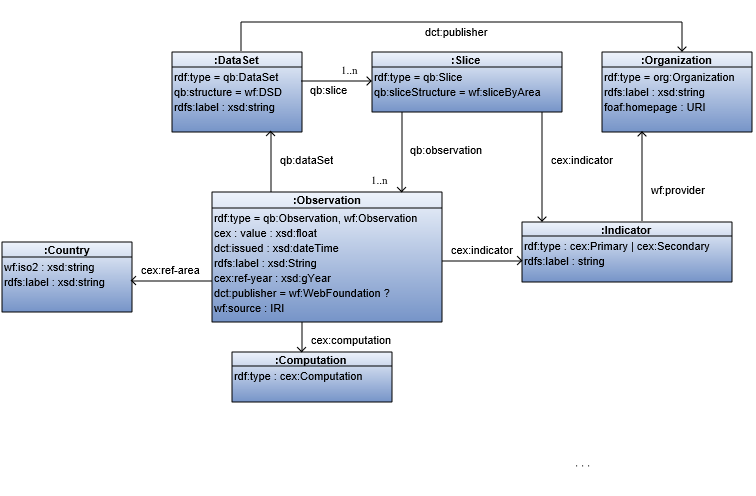}
\end{center}
\caption{Simplified WebIndex data model}
\label{SimplifiedModel}
\end{figure*}

The WebIndex data model is based on the RDF Data Cube vocabulary~\cite{RDFDataCube}
and reuses several vocabularies like the Organizations ontology~\cite{Organization14}, Dublin Core~\cite{DublinCore}, etc. Figure~\ref{SimplifiedModel} represents the main concepts of the data model\footnote{In the paper we will employ common prefixes (e.g. \lstinline|rdf|, \lstinline|foaf|, \lstinline|dct|, etc.) that can be found in \url{http://prefix.cc}. 
In addition, the \lstinline|wf| prefix represents the Web Foundation domain specific ontology}. 
The boxes represent the different shapes of nodes that are published in the data portal.

As can be seen, the main concept is an observation of type \lstinline|wf:Observation| which has a float value \lstinline|cex:value| for a given indicator, as well as the country, year, and dataset. 
Observations can be raw observations, which are obtained from an external source, or computed observations, 
which are obtained from other observations by some computation process. 

A dataset contains a number of slices, each of which also contains a number of observations. 

Indicators are provided by an organization of type \lstinline|org:Organization| which employs the Organization ontology\cite{Organization14}. Datasets are also published by organizations. 

A sample from the  DITU  dataset provided by ITU (International Telecommunication Union) states 
that, in 2011, Spain had a value of 23.78  for the TU-B (\emph{Broadband subscribers per 100 population}) indicator. This information is represented in RDF using Turtle syntax as:

\begin{lstlisting}[style=Turtle]
:obs8165 
 a qb:Observation, wf:Observation ;
 rdfs:label "ITU B in ESP" ;
 dct:issued
  "2013-05-30T09:15:00"^^xsd:dateTime ;
 cex:indicator   :ITU_B ;
 qb:dataSet      :DITU ;
 cex:value       23.78^^xsd:float ;
 cex:ref-area    :Spain ;
 cex:ref-year    2011 ;
 cex:computation :comp234 .
\end{lstlisting}

The WebIndex data model contains data that is completely interrelated. 
Observations are linked to indicators and datasets. Datasets contain also links to slices and slices have links to 
indicators and observations again. 
Both datasets and indicators are linked to the organizations that publish or provide them. 

The following example contains a sample of interrelated 
data for this domain. 
 
\begin{lstlisting}[style=Turtle]
:DITU a        qb:DataSet ;
 qb:structure  wf:DSD ;
 rdfs:label    "ITU Dataset" ;
 dct:publisher :ITU ;
 qb:slice      :ITU09B , 
               :ITU10B, 
               ...
:ITU09B a          qb:Slice ;
 qb:sliceStructure wf:sliceByArea ;
 qb:observation    :obs8165,
                   :obs8166,
                   ...
:ITU a         org:Organization ;
 rdfs:label    "ITU" ;
 foaf:homepage <http://www.itu.int/> .
 
:Spain 
 wf:iso2    "ES" ;
 rdfs:label "Spain" .
 
:ITU_B a     wf:SecondaryIndicator ;
 rdfs:label  "Broadband subscribers %";
 wf:provider :ITU . 
\end{lstlisting}

For verification, the WebIndex data model also includes a representation of computations that declares how each observation has been obtained, either from a raw dataset or computed from the observations of other datasets. We omit the description of computations in this paper for simplicity. That structure was presented in~\cite{LabraMTSR14}. 

In the next section we define formally the structure of this simplified WebIndex data model using ShEx 
and review the main differences with the original one.

\section{Using ShEx to describe the WebIndex data model}
\label{sec:ShapeExpressions}

ShEx has a compact syntax oriented towards human readability and can also be serialized in JSON, RDF and XML.  An introduction to ShEx can be found at \url{http://shex.io/primer/}.

ShEx uses the notion of a \lstinline[style=ShExC]|Shape| to describe RDF graph structures. A ShEx \lstinline[style=ShExC]|Shape|  describes the triples touching a given node in an RDF graph. 
Syntactically, it is a pairing of a label, which can be an IRI or a 
blank node, and a rule enclosed in brackets 
(\lstinline|{ }|).
A typical rule consists of a group of constraints separated 
by commas (\lstinline|,|) indicating that all the constraints must be satisfied.
For example, we can declare the shape of a country as:

\label{ShExCountry}
\begin{lstlisting}[style=ShExC]
:Country { 
 rdfs:label xsd:string,
 wf:iso2    xsd:string
}
\end{lstlisting}

The above declaration indicates that a valid \lstinline|:Country| shape must have exactly one
\lstinline|rdfs:label| and exactly one \lstinline|wf:iso2| both of which must be literals 
of type \lstinline|xsd:string|.

It should be noted that \lstinline|rdf:type| may or may not be included in shape definitions.  
In the above example, we deliberately 
omitted the \lstinline|rdf:type| requirement declaration, meaning that, in order to satisfy the \lstinline|:Country| shape, a node need only have the properties that we have specified. 
By default, shape definitions are ``open'', 
meaning that additional triples with different predicates may be present
so nodes of shape \lstinline|:Country| could have other properties apart 
of the properties which have been prescribed by its shape.

The ShEx language rules use the standard regular expression cardinality values of
\lstinline|+| (one or more), \lstinline|*| (zero or more),  
\lstinline|?| (zero or one) and 
\lstinline|{m,n}| (between \lstinline|m| and \lstinline|n| repetitions). The default rule cardinality is \lstinline|{1,1}| (exactly one).

Property values can be declared as sets of possible values \lstinline|[| .. \lstinline|]|\footnote{In previous versions of ShEx we used parenthesis instead of square brackets to represent value sets} or as value types (e.g. \lstinline|xsd:string|, \lstinline|IRI|). 
It is also possible to declare that the value of some property has a given shape using the \lstinline|@| character. 

For example, the shape of datasets can be described as:

\begin{lstlisting}[style=ShExC]
:DataSet { a   [ qb:DataSet ], 
 qb:structure  [ wf:DSD ], 
 rdfs:label    xsd:string ?, 
 qb:slice      @:Slice +, 
 dct:publisher @:Organization
} 
\end{lstlisting}

\noindent{}which declares that nodes satisfying \lstinline|:DataSet| shape must have an \lstinline|rdf:type| of \lstinline|qb:DataSet|, 
a \lstinline|qb:structure| of \lstinline|wf:DSD|, 
an optional \lstinline|rdfs:label|
of type \lstinline|xsd:string|, 
one or more \lstinline|qb:slice| predicates whose object is the subject of a set of triples 
matching the \lstinline|:Slice| shape definition and exactly one \lstinline|dct:publisher|, 
whose object is the subject of a set of triples matching the \lstinline|:Organization| shape.

The \lstinline|:Slice| shape is defined in a similar fashion:

\begin{lstlisting}[style=ShExC]
:Slice { a         [ qb:Slice ], 
 qb:sliceStructure [ wf:sliceByYear ],
 qb:observation    @:Observation+, 
 cex:indicator     @:Indicator
}
\end{lstlisting}

The \lstinline|:Observation| shape in the WebIndex data model has two \lstinline|rdf:type| declarations, which indicate that 
they must be instances of both the RDF Data Cube class of Observation (\lstinline|qb:Observation|) and the
\lstinline|wf:Observation| class from the Web Foundation ontology.
The property \lstinline|dct:publisher| is optional, but if it appears, it must have value \lstinline|wf:WebFoundation|.

Instances of the \lstinline|:Observation| shape can either have a \lstinline|wf:source| property of type \lstinline|IRI| (which, in this context, is used to indicate that it is a
raw observation that has been taken from the source represented by the IRI\footnote{It should be noted that ShEx do \emph{not} define the semantics of an RDF graph.  While the designers of the WebIndex dataset 
model have determined that a raw observation would be indicated using the \lstinline|wf:source| predicate and with the object \lstinline|IRI| referencing the original source, ShEx simply states that, in order for a 
subject to satisfy the \lstinline|:Observation|, it must include either a \lstinline|wf:source| or a \lstinline|cex:computation| predicate, period. Meaning must be found elsewhere.}) or a \lstinline|cex:computation| property whose object is the subject 
of a shape that satisfies the \lstinline|:Computation| constraint.

\begin{lstlisting}[style=ShExC]
:Observation { 
 a                 [ qb:Observation ], 
 a                 [ wi:Observation ], 
 cex:value         xsd:float, 
 dct:issued        xsd:dateTime, 
 dct:publisher     [wf:WebFoundation]?, 
 qb:dataSet        @:DataSet, 
 cex:ref-area      @:Country, 
 cex:indicator     @:Indicator, 
 cex:ref-year      xsd:gYear, 
 ( wf:source       IRI 
 | cex:computation @:Computation
 )
}
\end{lstlisting}

A computation is represented as a node with type \lstinline|cex:Computation|. 

\begin{lstlisting}[style=ShExC]
:Computation {
 a [ cex:Computation ]
}
\end{lstlisting}

Indicators are defined as:
 
\begin{lstlisting}[style=ShExC]
:Indicator {
 a       [ wf:PrimaryIndicator 
           wf:SecondaryIndicator ], 
 wf:provider @:Organization
}
\end{lstlisting}

In the case of organizations, we declare that they are closed shapes using the \lstinline|CLOSED| modifier so we only allow the properties \lstinline|rdfs:label|, \lstinline|foaf:homepage| and \lstinline|rdf:type|, which must have the value \lstinline|org:Organization|. 
The \lstinline|EXTRA| modifier is used to declare that we allow other values for the \lstinline|rdf:type| property (using the N3/Turtle keyword \lstinline|a|).

\begin{lstlisting}[style=ShExC]
:Organization CLOSED EXTRA a {
 a             [ org:Organization ], 
 rdfs:label    xsd:string, 
 foaf:homepage IRI
}
\end{lstlisting}

As can be seen, Shape Expressions offer an intuitive way to describe the contents of linked data portals. In fact, we have employed  Shape Expressions to document both the 
WebIndex\footnote{\url{http://weso.github.io/wiDoc}} and 
the Landbook\footnote{\url{http://weso.github.io/landportalDoc/data}} data portals.
The documentation defines templates for the different shapes of 
 resources and for the triples that can be retrieved when dereferencing
 those resources. 
 
These templates define the dataset structure in a declarative way and
 can be used to act as a contract between developers of the data portal. 
 We noted that having a good data model with its corresponding Shape Expressions specification facilitated the communication between the different stakeholders involved in the data portal development.

\paragraph{Differences with the original data model.} 
The data model described in this paper differs from
 the original one which was described at~\url{http://weso.github.io/wiDoc/}. The main differences are:
 \begin{itemize}
 \item Simplified model. We have omitted the representation of computations, which are
  represented as single nodes with type \lstinline|cex:Computation|. 
  A more detailed description of computations was described at~\cite{LabraMTSR14}.
  We have also simplified the representation of the webindex structure. 
  The original one was composed of sub-indexes and components. 
  Those features are easy to model and including them in this paper would 
  not offer any insight about the modelling expressiveness. 
  We also omitted several repeated properties like \lstinline|skos:notation|, \lstinline|rdfs:comment|, etc.
  
 \item Enriched representation of observations. In this we defined 
 observations with two \lstinline|rdf:type| declarations
 to indicate that there is a separation between classes and shapes. 
 We also included a disjunction to associate either computations or raw values to observations. 
 The original data model already had the separation between raw observations and computed observations. 
 However, at the time we defined the original model we were not sure how to represent them so we 
 represented raw observations as computations of type \lstinline|cex:Raw|. We consider that using 
 disjunction is a more natural way to represent them. 
 
 \item No mandatory \lstinline|rdf:type| arc for countries. 
  We define the shapes of countries to include just two simple properties.
  We deliberately omit the mandatory use of \lstinline|rdf:type| declaration to show that 
  it is possible to have nodes without that declaration. 
  In the original WebIndex data model, there were several nodes which did not 
  have \lstinline|rdf:type| declarations. 
  However, as in this paper we omit the representation of 
  computations we decided to offer that possibility for countries.  
  
\item CLOSED and EXTRA features for organizations. 
 We added those features to the \lstinline|:Organization| shape in order 
 to show their usage in this context. 
 \end{itemize}

\subsection{Shape Expressions Tools}
\label{ShExTools}

Currently, there are several implementations of 
ShEx in progress\footnote{More information about implementations can be found at \url{http://shex.io}}:

\begin{itemize}
\item \emph{ShEx.js}\footnote{\url{https://github.com/shexSpec/shex.js/}} is a Javascript implementation. 
 It handles semantic actions which can be used
 to extend the semantics of shape expressions and even to transform RDF to XML or JSON. 
 The Javascript code has been used by the \href{http://www.w3.org/2013/ShEx/FancyShExDemo}{ShExDemo}, a form-based system with 
 dynamic validation during the edition process and SPARQL queries generation.
 
\item ShExcala\footnote{\url{http://labra.github.io/ShExcala/}} is an implementation developed in Scala. 
It supports validation against an RDF file and against a SPARQL endpoint. 
Given that is implemented in Scala and it compiles to the JVM, the library can be called by Java 
or any other language that works on the JVM. We are currently adding support for ScalaJs, so it can 
also be compiled and run as a Javascript library.
 

\item Shexypy\footnote{\url{https://github.com/hsolbrig/shexypy}}: a Python implementation which contains an ANTLR4 parser and an interpreter. The shexypy implementation
uses XML as the primary representational form for Shape Expressions.  It uses the \lstinline[language=Python]|regex| package to evaluate shapes and \lstinline[language=Python]|rdflib|
for dataset access and query.
  
\item Haws\footnote{\url{http://labra.github.io/Haws/}}: a Haskell implementation based on type inference semantics and backtracking. 
This implementation was intended to be an executable monadic semantics of Shape Expressions~\cite{Labra2002}.

\item{RDFShape}\footnote{\url{http://rdfshape.weso.es}} is an online RDF Shape validation service that can be used to validate both the syntax and the shape of RDF data against some schema. 
The online service can get RDF data from an external URI, by file upload or manually written in a textarea field. 
It can also be used to validate against an external endpoint or by URI dereferentiation. 

RDFShape can be used as a web service which can be called using different query parameters as well as a 
simple Web application. Internally, it has been implemented using the Play! Framework.
It can also be configured to use either ShEx (using the ShExcala library) 
or SHACL (using either our own SHACL version or the TopQuadrant SHACL API engine). 

\end{itemize}


\subsection{Validating linked data portals using ShEx}
\label{sec:ValidatingWithShex}

ShEx can be used not only to describe the contents of linked data portals, but also to validate them.
We consider that one of the first steps in the development of a linked data portal 
should be the shapes declarations of the different resources which are published. 
ShEx can play a similar role to Schema declarations in XML based developments. 
They can act as a contract for both the producers and consumers of linked data portals.

Notice, however, that this contract does not suppose an extra-limitation between the 
possible consumers a linked data portal can have. 
There is no impediment to have more than one shape expression which enforce different constraints. 
As a na\"{\i}ve example, the declarations of the \lstinline|wf:iso2| code of countries can be 
further constrained using regular expressions to indicate that they must be 2 characters or 
could be more relaxed saying that it may be any value (not only strings).
The advantage of ShEx is that they offer a declarative and intuitive language to 
express and refer to those constraints.

ShEx can also be employed to generate synthetic linked data in the development phase so one can perform stress tests. 
For example, during the development of the WebIndex data portal, we implemented the \emph{wiGen}\footnote{\url{https://github.com/labra/wiGen}} tool which can generate random data that 
follows the WebIndex data model. 
These fake RDF datasets can be employed to perform stress and usability tests of the data visualization software.
In section~\ref{sec:PerfomanceComparison} we describe how we used \emph{wiGen} to perform some performance benchmarking.


Shexcala offers the possibility to validate resources from an endpoint or by dereferencing URIs. 
The validation through an endpoint performs a SPARQL query to obtain all the triples that have a given node as subject or object in the endpoint. 
Once the triples are retrieved, the system validates the ShEx declarations of that graph to check the shape of that node. 
In this way, it is possible to perform 
 shape checking on the contents of linked data portals. 
 
Notice that in general, this kind of validation is context sensitive to a given data portal. ShEx deliberately separates classes from shapes. 
We consider that shapes represent specifications about the structure of nodes in RDF graphs, while classes usually represent concepts from some domain. 
As an example, when we defined the data model of a similar data portal (the LandPortal) we also defined observations that were instances of \lstinline|qb:Observation| but had different shapes.
Both WebIndex and LandPortal respect the RDF data Cube definition of the class \lstinline|qb:Observation|, but they can use different properties (from that ontology or elsewhere) on those Observations, 
i.e. the observations in WebIndex have different shapes than the observations in LandPortal, 
but all of them have type \lstinline|qb:Observation| without introducing any logical conflicts.

We consider that differentiating structural shapes and the semantic types of resources improves the 
separation of concerns involved in linked data portal development. 
Nevertheless, although some shapes can be specific to some linked data portals, nothing precludes to 
define templates and libraries of generic shapes that can be reused between different data portals.
 
\section{Describing the Webindex using SHACL}
\label{sec:ValidatingWithSHACL}

\normalsize

In 2014, the W3C chartered the RDF Data Shapes Working Group to ``produce a language for defining structural constraints on RDF graphs\cite{SHACLCharter}''. The name chosen was SHACL (Shapes Constraint Language) and a First Public Working Draft (FPWD) was published in October 2015~\footnote{\url{http://www.w3.org/TR/shacl/}}.
In that version, SHACL is divided in two parts. The first part describes a core RDF vocabulary to define common shapes and constraints while the second part, titled "Advanced Features" describes an extension mechanism in terms of SPARQL queries and templates. 
SHACL is currently under development and there is not yet consensus inside the Working Group. In this section we will use the SHACL version published
in the FPWD and will concentrate on the first part, the core vocabulary.

SHACL groups the information and constraints that apply to a given data node into ``shapes''. 
A  SHACL \lstinline|sh:Shape| defines a collection of constraints that describe the structure of a given node. It may also include a ``scope definition'' that identifies the set of nodes to be tested for conformance.
A SHACL implementation interprets a collection of SHACL shape definitions against the scope nodes and determines whether the set of nodes conform to the definition. 

An equivalent SHACL description for the \lstinline|:Country| shape defined in page~\ref{ShExCountry} would be:

\begin{lstlisting}[style=Turtle]
:Country a sh:Shape ;
 sh:property [
  sh:predicate rdfs:label ;
  sh:datatype xsd:string ;
  sh:minCount 1; sh:maxCount 1 ;
] ;
 sh:property [
  sh:predicate wf:iso2 ;
  sh:datatype xsd:string ;
  sh:minCount 1; sh:maxCount 1 ;
] ;
\end{lstlisting}

As can be seen, the \lstinline|:Country| shape is defined by two constraints which specify that the datatype of \lstinline|rdfs:label| and \lstinline|wf:iso2| properties must be \lstinline|xsd:string|.

There are several ways to inform a SHACL implementation which nodes should be validated with which shapes, the simplest of which is by declaring that the scope node of a Shape is some given node using the
\lstinline|sh:scopeNode| predicate:

\begin{lstlisting}[style=Turtle]
:Country sh:scopeNode :Spain .
\end{lstlisting}

Another possibility is to associate a shape with every instance of
 some given class using the \lstinline|sh:scopeClass| predicate. 
 This approach can be used for what has been called ``record classes''~\cite{EricRecordClasses15}.

The default SHACL cardinality constraint is \lstinline|[0..*]| meaning that cardinality constraints that are omitted in the Shape Expressions grammar must be explicitly stated as:
\begin{lstlisting}[style=Turtle]
sh:minCount 1; sh:maxCount 1 ;
\end{lstlisting}
in SHACL.  Optionality (\lstinline|?| or \lstinline|*| in Shape Expressions) can be represented either by omitting \lstinline[style=Turtle]|sh:minCount| or by  \lstinline[style=Turtle]|sh:minCount=0|.  An unbounded maximum
cardinality (\lstinline|*| or \lstinline|+| in Shape Expressions) must be represented in SHACL by omitting  \lstinline[style=Turtle]|sh:maxCount|.
As an example, the definition of the \lstinline|:DataSet| shape declares that \lstinline|rdfs:label| is optional omitting the \lstinline|sh:minCount| property and declares that there must be one or more \lstinline|qb:slice| predicates conforming to the \lstinline|qb:slice| definition by omitting the value of \lstinline|sh:maxCount|.

The predicate \lstinline|sh:valueShape| is used to indicate that the value of a 
property must have a given shape. In this way, a shape can refer to another shape. 
It is possible that those shapes refer to other shapes and that these references 
form a cyclic data model as is the case of the WebIndex. Handling recursion in 
SHACL is an open issue in the current draft because it is not supported by SPARQL, 
the underlying technology on which SHACL is based.

\begin{lstlisting}[style=Turtle]
:DataSet a sh:Shape ;
 sh:property [
  sh:predicate rdf:type ;
  sh:hasValue qb:DataSet ;
  sh:minCount 1; sh:maxCount 1 ;
] ;
 sh:property [
  sh:predicate qb:structure ;
  sh:hasValue wf:DSD ;
  sh:minCount 1; sh:maxCount 1 ;
] ;
 sh:property [
  sh:predicate rdfs:label ;
  sh:datatype xsd:string ;
  sh:maxCount 1 ;
] ;
 sh:property [
  sh:predicate qb:slice ;
  sh:valueShape :Slice ;
  sh:minCount 1 ;
] ;
 sh:property [
  sh:predicate dct:publisher ;
  sh:valueShape :Organization ;
  sh:minCount 1; sh:maxCount 1 ;
] .
\end{lstlisting}

The definition of \lstinline|:Slice| is similar to \lstinline|:DataSet| so we can omit it for clarity. 
The full version of the SHACL shapes that we used in the paper are available at the wiGen repository\footnote{\url{https://github.com/labra/wiGen/blob/master/schemas/webindexShapes.ttl}}.

There are three items that need more explanation in the SHACL definition of the \lstinline|:Observation| shape. 
The first one is the repeated appearance of the \lstinline|rdf:type| property with two values. 
Although we initially represented it using qualified value shapes, we noticed that it could also be 
represented as: 

\begin{lstlisting}[style=Turtle]
:Observation a sh:Shape ;
 sh:property [
  sh:predicate rdf:type ;
  sh:in ( qb:Observation 
          wf:Observation ) 
 sh:property [
  sh:predicate rdf:type ;
  sh:minCount 2; sh:maxCount 2
] ;
...
\end{lstlisting}

The definition of observations also contains an optional property with a fixed value which was defined in ShEx as:

\begin{lstlisting}[style=ShExC]
:Observation {  ...
 dct:publisher (wf:WebFoundation)? 
 ...
}
\end{lstlisting}

\noindent{}which means that observations can either have a property \lstinline|dct:publisher| with the fixed 
 value \lstinline|wf:WebFoundation| or not have that property. 

A first approach to model that in SHACL would be to use \lstinline|sh:minCount| with cardinality 0 but that declaration 
contradicts \lstinline|sh:hasValue| so it is necessary to use \lstinline|sh:filterShape| to indicate that the 
constraint is only applied to nodes that have the \lstinline|dct:publisher| property.

\begin{lstlisting}[style=Turtle]
:Observation  ...
 sh:property [
  sh:predicate dct:publisher ;
  sh:hasValue wf:WebFoundation ;
  sh:filterShape [
   sh:property [
    sh:predicate dct:publisher ;
    sh:minCount 1 ;
  ]] ;        
 sh:maxCount 1 ;
 ] ; ...
\end{lstlisting}

The last item requiring additional explanation is the disjunction definition which indicates that observations 
must have either the property \lstinline|cex:computation| with a value of shape \lstinline|:Computation| or the property \lstinline|wf:source| with an IRI value, but not both. In ShEx, it was defined as:

\begin{lstlisting}[style=ShExC]
:Observation { ...
 , ( cex:computation @:Computation
   | wf:source IRI
   )
 ...
}
\end{lstlisting}

In SHACL, although there is a predefined \lstinline|sh:OrConstraint|, it is not exclusive, so it is necessary to impose 
 another constraint that forbids both to appear.

\label{DisjointOrShacl}
\begin{lstlisting}[style=Turtle]
:Observation 
 ...
 sh:constraint [
 a sh:OrConstraint ;
 sh:shapes (
 [ sh:property [
   sh:predicate wf:source ;
   sh:nodeKind sh:IRI ;
   sh:minCount 1; sh:maxCount 1 ;
 ]]
 [ sh:property [
   sh:predicate cex:computation ;
   sh:valueShape :Computation ;
   sh:minCount 1; sh:maxCount 1 ;
 ]])] ;
 sh:constraint [
 a sh:NotConstraint ;
 sh:shape [ 
 sh:constraint [
 a sh:AndConstraint ;
 sh:shapes (
 [ sh:property [
    sh:predicate wf:source ;
 	sh:nodeKind sh:IRI ;
    sh:minCount 1; sh:maxCount 1 
  ]]
  [ sh:property [
    sh:predicate cex:computation ;
    sh:valueShape :Computation ;
    sh:minCount 1; sh:maxCount 1 ;
 ]])]]]...
\end{lstlisting}
 
In the case of indicators we can see again the separation between \lstinline|:Indicator| shape and 
 the \lstinline|wf:PrimaryIndicator| and \lstinline|wf:SecondaryIndicator| classes.

\begin{lstlisting}[style=Turtle]
:Indicator a sh:Shape ;
 sh:property [
  sh:predicate rdf:type ;
  sh:in (
     wf:PrimaryIndicator 
     wf:SecondaryIndicator
  ) ;
  sh:minCount 1; sh:maxCount 1 ;
 ] ; 
 ...
\end{lstlisting}

Finally, we defined organizations as closed shapes with the possibility that the \lstinline|rdf:type| property had some extra values apart from the \lstinline|org:Organization|. This constraint can be expressed in SHACL as:

\begin{lstlisting}[style=Turtle]
:Organization a sh:Shape ;
  sh:constraint [
   a sh:ClosedShapeConstraint;
   sh:ignoredProperties(rdf:type)
  ] ;
 sh:property [
  sh:predicate rdf:type ;
  sh:hasValue org:Organization ;
 ] ;
 ...
\end{lstlisting}


\subsection{SHACL tools}
\label{subsec:SHACLTools}

\begin{itemize}
\item TopBraid SHACL API\footnote{\url{https://github.com/TopQuadrant/shacl}} 
is an implementation developed in Java using the Jena Library which is going to be 
used in the TopBraid products. This implementation supports recursive shapes 
and is the one that has been employed in this paper.
 
\item SHACL Engine\footnote{\url{https://github.com/x-lin/shacl-engine}} was another 
implementation developed in Java using the Jena Library. 
The SHACL engine was part of  SHACL4P~\footnote{\url{https://github.com/fekaputra/shacl-plugin}}, 
a SHACL plugin for Prot{\'e}g{\'e}.
The README of the project says that the implementation is currently deprecated.

\item shacl\footnote{\url{https://github.com/pfps/shacl}}, an implementation of SHACL developed by 
 Peter F. Patel Schneider in Python as a translation to SPARQL.
 
\item RDFUnit\footnote{\url{http://rdfunit.aksw.org/}} a test driven data-debugging framework has recently 
 added support for SHACL. 
 
\item Shaclex\footnote{\url{http://labra.github.io/shaclex}}. An experimental 
 implementation of SHACL combining aspects from ShEx. 
 It is part of the online RDFShape tool described in the previous section.
\end{itemize}
\section{Differences between ShEx and SHACL}
\label{sec:ComparingShExSHACL}

Both ShEx and SHACl can be used to describe the WebIndex data portal contents and 
their core features are similar. 
However, there are several differences between both formalisms like:

\begin{itemize}

\item Syntax. ShEx was designed following a grammar based approach which defined an abstract syntax and 
 its corresponding serializations.
 In this way, it is possible to separate the language from its syntax and we have already proposed 
 several concrete syntaxes as the compact syntax presented in this paper 
 and JSON and RDF serializations. 
 In the case of SHACL, the working group has opted to define using an RDF vocabulary, so 
 at the time of this writing there is only one RDF based serialization. 
 It was proposed to define also a user-friendly compact syntax for SHACL inspired by the one but 
 it is still work-in-progress. 
 
\item Negation and groupings. 
ShEx allows to define negations using the 
operator \lstinline$!$ and 
groupings using parenthesis that can express more complex patterns. 
For example, it is possible to declare that countries must have \lstinline|wf:iso2| and \lstinline|wf:iso3| at the same time, 
but that they may be optional, and that they 
must not have the property \lstinline|dc:creator| with any value (represented by dot in ShEx) as:

\begin{lstlisting}[style=ShExC]
:Country { a [ wf:Country ], 
 rdfs:label   xsd:string, 
 ( wf:iso2    xsd:string 
 | wf:iso3    xsd:string 
 )?, 
 ! dc:creator . 
}
\end{lstlisting}

In the case of SHACL, there is also a \lstinline|sh:NotConstraint|, 
as well as the \lstinline|sh:AndConstraint| for conjunction and \lstinline|sh:OrConstraint| 
for disjunction. 
In principle, it is not possible to group those operators and assign cardinalities 
to the resulting groups. 
However, although the ShEx formalism allows that kind of cardinality nesting, 
we are considering to restrict that expressiveness in order to avoid 
the complexity overload that they impose.

\item Shape inclusion. 
In ShEx, it is possible to reuse shape descriptions 
by including other shape declarations. For example, one may be interested to say that 
providers have the shape \lstinline|:Organization| but also contain the property \lstinline|wf:sourceURI| as:

\begin{lstlisting}[style=ShExC]
<Provider> & <Organization> { 
  wf:sourceURI IRI 
}
\end{lstlisting}

The SHACL working draft contains a section about templates and user-defined functions 
 which are expected to handle those cases.

\item Incoming edges. Although the examples presented in the previous section are based on describing the subjects of RDF graphs, 
it is possible to handle also objects. 
For example, one can declare reverse arcs using the \lstinline|^| operator to indicate incoming arcs. 
The Country declaration could be:

\begin{lstlisting}[style=ShExC]
:Country { 
 rdfs:label     xsd:string, 
 wf:iso2        xsd:string, 
 ^ cex:ref-area @:Observation *
}
\end{lstlisting}

with the meaning that a country can receive (zero or more) arcs with property \lstinline|cex:ref-area| of 
shape \lstinline|:Observation|. 
In the SHACL draft, there is a whole section about inverse properties that is expected to cover incoming edges.

\item Extension mechanism. 
ShEx can be extended with a feature called semantic actions to increase the expressiveness\footnote{The name semantic actions is inspired by parser generators but it is not related to any kind of semantics as employed in the context of semantic web}. 
Semantic actions are marked by \lstinline|%lang {  actions %}| which means that the validator can invoke a processor of the language \lstinline|lang| with the corresponding actions. 
The JavaScript implementation supports semantic actions in JavaScript and SPARQL which can add more expressiveness to the validation declarations. It also contains two simple languages (GenX and GenJ) which enable an easy way to transform RDF to both XML and JSON. 

In the case of SHACL, the second part of the draft is devoted to \emph{Advanced Features} and contains native constraints 
defined in SPARQL, templates, scopes and even functions. 

%
% ...
%
%
%

\item Selection of nodes to validate. 
 The selection of which nodes are going to be selected for validation
 has been let unspecified in ShEx specification, 
 which lets that choice to the validation engine. 
 SHACL defines several options using the concept of scopes. 
 Scopes can be individual scopes which associate a shape with a single node, 
 class scopes, with associate a shape with all the instances of some class, or general 
 scopes, which offer a more generic mechanism to select focus nodes.

\item Reasoning. The interplay between reasoners and ShEx or SHACL is not established. 
Some applications could do validation on RDF graphs that include entailments, which could be 
pre-computed before validation of computed on the fly during validation. 
SHACL mentions the property \lstinline|sh:entailment| to 
instruct a SHACL validation engine that a given entailment should be activated
but SHACL processors are not required to support entailment regimes.

\item Recursion. ShEx allows recursion to define cyclic definitions of data models. 
 The SHACL draft does not allow recursive shapes and 
 the behaviour of the implementations is undefined. 
 The TopBraid SHACL implementation allows recursive 
 shapes and handles the WebIndex data model for small graphs.
 In the case of the WebIndex data model, if we impose the constraint that every node 
 has an \lstinline|rdf:type| arc indicating the class to which it belongs, 
 then it is possible to describe the whole model without recursion by 
 adding the property \lstinline|sh:scopeClass| to associate each shape with the corresponding class.
 We have employed two possible shapes definitions for the webindex data portal, 
 one with recursive shapes and one with non-recursive shapes~\footnote{The schemas are available at~\url{http://labra.github.io/wiGen/}}. 

\item Underlying formalism. 
ShEx is based on a generalization of regular expressions to handle unordered sets called 
 regular bag expressions. 
ShEx has a well defined semantics that covers recursion. 
The SHACL core is based on the notion of shapes which group constraints and their constructs are defined in 
terms of SPARQL. 
We are currently working on providing a more formal comparison between 
ShEx and SHACL expressivity~\cite{boneva:hal-01288285}
where we define a generic Shapes constraint language and show we can translate SHACL core constructs 
to it so it will be possible to implement SHACL by conversion to ShEx.
\end{itemize}

\section{Performance Benchmarking Tool}
\label{sec:PerfomanceComparison}

\begin{table*}[!h]
\caption{Elapsed time (ms) to validate all nodes using recursive shapes}
\label{tab:resultsAll}
\begin{center}
\begin{tabular}{ l r r r r r r r }
\toprule
 & \multicolumn{7}{c}{Value} \\
Parameter & 1 & 5 & 10 & 50 & 100 & 500 & 1000 \\
\hline
Countries  &  135 &  145 &  161 & 315 &  493 &  3838 &  21445 \\
Datasets & 133 & 157 & 189 & 445 & 716 & 7071 & 41676 \\
Slices & 135 & 159 & 188 & 448 & 800 & 13689 & 93280 \\
Observations &  136 &  229 & 357  & 967 & 1738 & 45504 & 331480 \\
Computations & 136 & 144 & 154 & 273 & 433 & 3943 & 23482 \\
Indicators & 135 & 155 & 173 & 415 & 645 & 7074 & 43699 \\
Organizations &  136 &  152 &  176 &  384 &  590 & 4135 & 22082 \\
\bottomrule\\
\end{tabular}
\end{center}
\end{table*}

The performance of shape checking directly limits the workflows that may exploit it, as well as users' general willingness to test data.
With regards to ShEx, \cite{Slawek2015} determined that the general complexity of Shape Expressions for open and closed shapes without negation is NP-complete. 
This paper isolated tractable fragments
and proposed a validation algorithm for a restriction called deterministic single-occurrence shape expressions and \cite{BonevaShapeExpressionSchemas15} presented a well founded semantics of Shape Expressions with negation and recursion as well as a full validation algorithm and guidelines for its efficient implementation. 
The SHACL draft is defined in terms of SPARQL, so its complexity depends in part on SPARQL's complexity~\cite{Perez09,Arenas12}, though the emerging algorithms for iterating across nodes and shapes are likely to have a larger effect.

Apart from the theoretical complexity, it is important to check the performance in practice.  
We have created a performance evaluation tool, 
\emph{wiGen}, that generates random WebIndex data that can be used as a benchmark\footnote{The tool is available at~\url{https://github.com/labra/wiGen}}. 
It takes as parameters the number of desired countries, datasets, slices, observations, indicators and organizations and generates valid RDF data according to the ShEx and SHACL schemas presented in previous sections. 
It can also be configured to generate a given number of not valid nodes of the different shapes and to
 add scope node declarations for all or for only one node.
The \emph{wiGen} tool can then be used to show the elapsed time needed to validate the generated data against different ShEx or SHACL implementations. 
At this moment, it can be configured to use either a ShEx implementation (ShExcala) 
or a SHACL one (TopQuadran SHACL API). 
Our first experiments compared performance between both implementations. 
However, upon request from the author we are not including results from the SHACL implementation 
in this paper because he said that it is a proof-of-concept implementation
which is not optimized for performance.

In our evaluation we employed an Intel Core 2 Duo CPU at 2.93GHz (3Gb RAM) using Debian Linux.  
In our first experiment, we check the validity of datasets generated with valid random values
 and changing the number of nodes of each different shape. 
The results can be seen in figure~\ref{tab:resultsAll} where we 
included the elapsed time in milliseconds.
The results show that the implementation's calculation time grows 
considerably when increasing the number of shapes. 
We also tried to validate the data with more realistic parameters that resemble 
the number of nodes that were available in the WebIndex data portal. 
When we run the experiment with 
80 countries, 
40 datasets, 
80 slices, 
5000 observations, 
4000 computations, 
50 indicators and 
4 organizations, 
the time was 84793079 milliseconds (23.55 hours).
Given that the implementation is still proof-of-concept, further work must 
be done in terms of profiling and performance optimization to identify which 
parts can be improved.


The \emph{wiGen} tool also contains a set of parameters to generate a given number 
invalid nodes of each of the shapes. 
In this way, it can be used to measure the time it takes the validator to check datasets
with invalid data.
Our next experiment was to execute the tool varying the number of shapes as in the previous one, 
but generating a single invalid node in each shape. 
The results are shown in figure~\ref{tab:resultsInvalid}. As can be seen, the 
implementation takes less time to identify an invalid dataset than to verify that 
all the nodes are valid. 

\begin{table*}[!h]
\caption{Elapsed time (ms) to check that a node is not valid}
\label{tab:resultsInvalid}
\begin{center}
\begin{tabular}{ l r r r r r r r }
\toprule
  & \multicolumn{7}{c}{Value} \\
Parameter     & 1   &   5 &  10 &  50 & 100 & 500 & 1000 \\
\hline
Countries     & 128 & 146 & 157 & 250 & 392 & 1895 & 8685 \\
Datasets      & 86 & 88 & 88 & 93 & 100 & 144 & 313 \\
Slices        & 104 & 119 & 136 & 300 & 466 & 2446 & 8406 \\
Observations  & 125 & 222 & 324 & 808 & 1325 & 18788 & 133012 \\
Computations  & 137 & 151 & 144 & 181 & 279 & 1246 & 4897 \\
Indicators    & 134 & 154 & 174 & 311 & 465 & 2445 & 11613 \\
Organizations & 96 & 141 & 179 & 372 & 565 & 3732 & 19554 \\
\bottomrule\\
\end{tabular}
\end{center}
\end{table*}

Another aspect that can affect validation time is the use of recursive shapes. 
 We have added a parameter to \emph{wiGen} to declare the class to which 
 each node belongs. In this way, it is possible to validate the data 
 without using recursive shapes. 
 Figure~\ref{tab:resultsNoRecursion} shows the timings when using 
 non-recursive shapes. 
 As can be seen, the times are smaller than
 when using recursive shapes, although the timings still increase when 
 the number of shapes is bigger. 
 In fact, we also tried with the previous pattern of 80 countries, 40 datasets, 80 slices, etc. and 
 the elapsed time was 76128579ms (21.14 hours). 
 As can be seen the validation time taken by the ShEx implementation 
 doesn't seem to be affected so much by the recursive/non-recursive 
 nature of the shapes. 

As can be seen, the \emph{wiGen} tool can be scripted to explore many relevant parameters: 
size of the validation graph, 
number of nodes to be validated, 
interrelations between nodes in recursive shapes and 
number of invalid nodes.
The use of this kind of tools to generate benchmark data will permit principled design choices in language development 
and tool selection, 
and ultimately contribute to improved quality in linked data.

The take-home message from this \emph{very} preliminary evaluation is that, 
while the performance figures still leave much to be desired, 
reasonable performance is definitely reachable.

Another aspect to take into account is that this approach to validate linked data portals is not 
realistic for very big datasets and it may be necessary to develop other algorithms which validate 
nodes behind endpoints on demand with some caching capabilities to avoid repeating the validation 
on already checked nodes. 

\begin{table*}[!h]
\caption{Elapsed time (ms) to validate all nodes using non-recursive shapes}
\label{tab:resultsNoRecursion}
\begin{center}
\begin{tabular}{ l r r r r r r r }
\toprule
 & \multicolumn{7}{c}{Value} \\
Parameter & 1 & 5 & 10 & 50 & 100 & 500 & 1000 \\
\hline
Countries     &  151 &  167 &  188 & 370 &  550 &  4078 &  21858 \\
Datasets      & 149 & 180 & 205 & 463 & 784 & 7161 & 41866 \\
Slices        & 152 & 190 & 228 & 583 & 1057 & 17072 & 109421 \\
Observations  & 150 &  243 & 364  & 978 & 1598 & 24688 & 173815 \\
Computations  & 148 & 159 & 167 & 283 & 431 & 3984 & 23625 \\
Indicators    & 152 & 167 & 192 & 383 & 601 & 4401 & 23023 \\
Organizations &  158 & 171 &  188 & 391 & 586 & 4177 & 22222 \\
\bottomrule\\
\end{tabular}
\end{center}
\end{table*}


\section{Related work}
\label{sec:Related}
\normalsize

Improving the quality of linked data has been of increasing interest in the last years. 
Zaveri et al~\cite{Zaveri2016} include a systematic survey on the subject. 
In that paper, they propose 18 quality dimensions for linked data quality like syntactic 
validity and semantic accuracy which will directly benefit if shapes are employed during 
the linked data portal development. 

In the case of RDF validation, the main approaches can be summarized as:

\begin{itemize}
\item Inference based approaches, which adapt RDF Schema or OWL to express validation semantics. 
The use of Open World and Non-unique name assumption limits the
validation possibilities. 
In fact, what triggers constraint violations in closed
world systems leads to new inferences in standard OWL systems.
\cite{Motik07} proposed the notion of \emph{extended description logics} knowledge bases, in which a certain subset of TBox axioms were designated as constraints. 
Peter F. Patel Scheneider separates the validation problem in two parts: 
 integrity constraint and closed-world recognition~\cite{Patel-Schneider2015}. 
He shows that 
 description logics can be implemented for both by translation to SPARQL queries. 
In 2010, Tao et al~\cite{Tiao10} had already proposed the use of OWL expressions with 
Closed World Assumption and a weak variant of Unique Name Assumption to express integrity constraints. 
Their work forms the bases of Stardog ICV~\cite{ClarkSirin13}, which is part of the Stardog database. 
It allows to write constraints in OWL and converts them to SPARQL queries. 
As an example, the country shape could be specified as:

\label{StardogCountry}
\begin{lstlisting}[style=ShExC]
:Country a owl:Class ;
 rdfs:subClassOf
  [ owl:onProperty rdfs:label ; 
    owl:minCardinality 1 ] ,
  [ owl:onProperty rdfs:label ; 
    owl:maxCardinality 1 ] ,
  [ owl:onProperty wf:iso2 ; 
    owl:minCardinality 1 ] ,
  [ owl:onProperty wf:iso2 ; 
    owl:maxCardinality 1 ] .

rdfs:label a owl:DatatypeProperty; 
  rdfs:range xsd:string .
  
wf:iso2 a owl:DatatypeProperty; 
  rdfs:range xsd:string .
\end{lstlisting}

It defines a class \lstinline|:Country| so instance nodes are supposed to 
 have an \lstinline|rdf:type| whose value should be that class in order to 
 be validated. 
 This is different from ShEx where shapes and classes don't 
 need to related.

\item SPIN. SPARQL Inferencing Notation (SPIN)\cite{SPIN11} was introduced by TopQuadrant as a mechanism 
to attach SPARQL-based constraints and rules to classes.  
SPIN also contained templates, user-defined functions and template libraries.
SPIN rules are expressed as SPARQL ASK queries where \lstinline|true| indicates an error or 
CONSTRUCT queries which produce violations. 
SPIN uses the expressiveness of SPARQL plus the semantics of the \lstinline|?this| variable standing for the current subject and violation class. 

\item  SPARQL-based approaches use the SPARQL Query 
Language to express the validation constraints. 
SPARQL has much more expressiveness than ShEx and can even be used to validate numerical and statistical computations. 
In fact, our first approach to validate the WebIndex data was to use SPARQL~\cite{Labra13}. 
However, we consider that ShEx will be more usable by people familiar with validation languages like RelaxNG. 

There have been other proposals using SPARQL combined with other technologies. 
F\"{u}rber and Hepp\cite{FurberH10} proposed a combination between SPARQL and SPIN as a semantic data quality framework,
Simister and Brickley\cite{Simister13} propose a combination
between SPARQL queries and property paths which is used by Google
and Kontokostas et al~\cite{kontokostasDatabugger} proposed \emph{RDFUnit} a Test-driven framework which employs SPARQL query templates that are
instantiated into concrete quality test queries. 
We consider that ShEx can also be employed in the same scenarios as SPARQL
while the specialized validation nature of ShEx can lead to more efficient
implementations. 

\item Grammar based approaches define a domain specific language to declare the validation rules. 
OSLC Resource Shapes~\cite{OSLCResourceShapes} 
have been proposed as a high level and declarative description of the expected contents of an RDF graph expressing constraints
on RDF terms. 
Dublin Core Application Profiles~\cite{KarenCoyleTomBaker13} also define a set of validation constraints using Description Templates with less expressiveness than Shape Expressions. 
Fischer et al~\cite{FischerEtAl:EDBT/ICDT-WS2015} proposed RDF Data Descriptions as a another domain specific language that is 
compiled to SPARQL. The validation is class based in the sense that RDF nodes are validated against a class \lstinline|C| 
whenever they contain an \lstinline|rdf:type C| declaration. This restriction enables the authors to handle the validation of large datasets and to define some optimization techniques which could be applied to shape implementations.
\end{itemize}

ShEx is as another grammar based approach which defines
 a domain specific language for RDF validation. 
 The grammar of ShEx was inspired by Turtle and RelaxNG focusing 
 in a target audience which could be familiar with these technologies. 
 In practice, we have found that users find the syntax intuitive.
 In fact, we employed ShEx to communicate the structure of triples that had to be generated
 to the WebIndex team of developers, which was comprised by people 
 familiar with Java, XML and some basic RDF, and they found it quite easy to understand ShEx.
 We also employed ShEx to document the data portal so consumers could understand the nodes 
 that were available. 
 
SHACL design has been mainly influenced by the SPIN approach. 
The Working Group has decided to offer a SPARQL based semantics and the second part of the 
working draft also contains the SPIN mechanism of 
SPARQL native constraints, templates and used-defined functions. 
There main differences are the renaming of some terms and the addition of more core constraints 
like disjunction, negation or closed shapes. 
At the time of this writing there are still several open issues about 
the inclusion of other features for SHACL.
The working group has also proposed the use of a concise syntax similar to ShEx for SHACL.
\section{Conclusions}
\label{sec:Conclusions}

The tools and techniques needed for linked data publishing are gradually maturing. However, there is yet a lack of tools to measure and guarantee the quality of linked data solutions. 
In fact, the medium of any linked data portal, RDF, still lacks a standard way to be described and validated.
Two approaches with the word \emph{shapes} in their acronym: ShEx and SHACL, have been proposed to validate the structure of RDF graphs.

ShEx was designed as a high-level and human-friendly notation based on a well founded semantics which can be implemented without the need of SPARQL processors~\cite{BonevaShapeExpressionSchemas15} while the First Public Working draft of SHACL has been defined as an RDF vocabulary whose semantics is defined in terms of SPARQL and some extensions to handle recursion and template invocation. 

In this paper we used both ShEx and SHACL to describe the contents of a medium-sized linked data portal. 
We consider ShEx to be more concise and intuitive than SHACL. 
Although this situation could be improved if a ShEx-like syntax were defined for SHACL, there remain some foundational differences between both languages that should be tackled, e.g. the combination between recursion, negation, grouping cardinalities, etc. which are not yet handled by the SHACL draft.

We created a generator of valid data according to the WebIndex schema of any size which can be used for benchmarking different validation engines and we measured the elapsed time of two different implementations of ShEx and SHACL.
This demonstrated how the \emph{wiGen} program can provide a detailed view of performance with respect to different axes.
This can, in turn, be used to profile tools and algorithms, informing the standardization of shapes languages, as well as enabling consumers to make informed tool choices.

Although these performance results are preliminary, we consider that separating the SHACL language from any particular technology, like the SPARQL templates implementation, could promote the development of new algorithms that may improve our current results. 
For instance, optimizing remote validation will be enhanced by compiled, monolithic SPARQL queries but that may make it difficult to caching the shape assignments performed by such a query.

In general we consider that the benefits of validation using either ShEx or SHACL can help the adoption of RDF based solutions where the quality of data is an important issue. The current work developed by the W3c Data Shapes Working group and the Shape Expressions community may help to improve RDF adoption in industrial scenarios where there is a real need to ensure the structure of RDF data, both in production and consumption. 


\bibliographystyle{abbrv}
\bibliography{ShEx}

\begin{thebibliography}{10}

\bibitem{Arenas12}
M.~Arenas, S.~Conca, and J.~P{\'e}rez.
\newblock Counting beyond a yottabyte, or how sparql 1.1 property paths will
  prevent adoption of the standard.
\newblock In {\em Proceedings of the 21st International Conference on World
  Wide Web}, WWW '12, pages 629--638, New York, NY, USA, 2012. ACM.

\bibitem{SHACLCharter}
C.~Arnaud Le~Hors.
\newblock {RDF Data Shapes Working Group Charter}.
\newblock \url{http://www.w3.org/2014/data-shapes/charter}, 2014.

\bibitem{bernerslee06}
T.~{Berners-Lee}.
\newblock Linked-data design issues.
\newblock W3C design issue document, June 2006.
\newblock http://www.w3.org/DesignIssue/LinkedData.html.

\bibitem{bizer_linked_2009}
C.~Bizer, T.~Heath, and T.~{Berners-Lee}.
\newblock Linked data - the story so far.
\newblock {\em International Journal Semantic Web Information Systems},
  5(3):1--22, 2009.

\bibitem{DublinCore}
D.~U. Board.
\newblock {DCMI Metadata Terms}.
\newblock \url{http://dublincore.org/documents/dcmi-terms/}, 2012.

\bibitem{boneva:hal-01288285}
I.~Boneva.
\newblock {Comparative expressiveness of ShEx and SHACL (Early working draft)}.
\newblock working paper or preprint, Mar. 2016.

\bibitem{BonevaShapeExpressionSchemas15}
I.~Boneva, J.~E.~L. Gayo, E.~G. Prud'hommeaux, and S.~Staworko.
\newblock Shape expressions schemas.
\newblock {\em CoRR}, abs/1510.05555, 2015.

\bibitem{ClarkSirin13}
K.~Clark and E.~Sirin.
\newblock On {RDF} validation, stardog {ICV}, and assorted remarks.
\newblock In {\em RDF Validation Workshop. Practical Assurances for Quality RDF
  Data}, Cambridge, Ma, Boston, September 2013. W3c,
  \url{http://www.w3.org/2012/12/rdf-val}.

\bibitem{KarenCoyleTomBaker13}
K.~Coyle and T.~Baker.
\newblock Dublin core application profiles. separating validation from
  semantics.
\newblock In {\em RDF Validation Workshop. Practical Assurances for Quality RDF
  Data}, Cambridge, Ma, Boston, September 2013. W3c,
  \url{http://www.w3.org/2012/12/rdf-val}.

\bibitem{RDFDataCube}
R.~Cyganiak and D.~Reynolds.
\newblock {The RDF Data Cube Vocabulary}.
\newblock \url{https://www.w3.org/TR/vocab-data-cube/}, 2014.

\bibitem{EDBT/ICDT-WS2015}
P.~M. Fischer, G.~Alonso, M.~Arenas, and F.~Geerts, editors.
\newblock {\em Proceedings of the Proceedings of the Workshops of the EDBT/ICDT
  2015 Joint Conference (EDBT/ICDT 2015)}, number 1330 in CEUR Workshop
  Proceedings, Aachen, 2015.

\bibitem{FischerEtAl:EDBT/ICDT-WS2015}
P.~M. Fischer, G.~Lausen, A.~Sch{\"a}tzle, and M.~Schmidt.
\newblock Rdf constraint checking.
\newblock In Fischer et~al. \cite{EDBT/ICDT-WS2015}, pages 205--212.

\bibitem{FurberH10}
C.~F\"{u}rber and M.~Hepp.
\newblock Using sparql and spin for data quality management on the semantic
  web.
\newblock In W.~Abramowicz and R.~Tolksdorf, editors, {\em Business Information
  Systems}, volume~47 of {\em Lecture Notes in Business Information
  Processing}, pages 35--46. Springer, 2010.

\bibitem{LabraMTSR14}
J.~E.~L. Gayo, H.~Farham, J.~C. Fern{\'{a}}ndez, and J.~M.~{\'{A}}.
  Rodr{\'{\i}}guez.
\newblock Representing statistical indexes as linked data including metadata
  about their computation process.
\newblock In S.~Closs, R.~Studer, E.~Garoufallou, and M.~Sicilia, editors, {\em
  Metadata and Semantics Research - 8th Research Conference, {MTSR} 2014,
  Karlsruhe, Germany, November 27-29, 2014. Proceedings}, volume 478 of {\em
  Communications in Computer and Information Science}, pages 42--53. Springer,
  2014.

\bibitem{LabraShexDerivatives15}
J.~E.~L. Gayo, E.~Prud'hommeaux, I.~Boneva, S.~Staworko, H.~Solbrig, and
  S.~Hym.
\newblock Towards an rdf validation language based on regular expression
  derivatives.
\newblock In Fischer et~al. \cite{EDBT/ICDT-WS2015}, pages 197--204.

\bibitem{SPIN11}
H.~Knublauch.
\newblock {SPIN - Modeling Vocabulary}.
\newblock \url{http://www.w3.org/Submission/spin-modeling/}, 2011.

\bibitem{kontokostasDatabugger}
D.~Kontokostas, P.~Westphal, S.~Auer, S.~Hellmann, J.~Lehmann, R.~Cornelissen,
  and A.~Zaveri.
\newblock Test-driven evaluation of linked data quality.
\newblock In {\em Proceedings of the 23rd International Conference on World
  Wide Web}, WWW '14, pages 747--758, Republic and Canton of Geneva,
  Switzerland, 2014. International World Wide Web Conferences Steering
  Committee.

\bibitem{Labra13}
J.~E. Labra and J.~M. Alvarez~Rodr\'{i}guez.
\newblock Validating statistical index data represented in {RDF} using {SPARQL}
  queries.
\newblock In {\em RDF Validation Workshop. Practical Assurances for Quality RDF
  Data}, Cambridge, Ma, Boston, September 2013. W3c,
  \url{http://www.w3.org/2012/12/rdf-val}.

\bibitem{Labra2002}
J.~E. Labra~Gayo.
\newblock Reusable semantic specifications of programming languages.
\newblock In {\em 6th Brazilian Symposium on Programming Languages}, 2002.

\bibitem{LabraLDQSemantics}
J.~E. Labra~Gayo, E.~Prud'hommeaux, H.~R. Solbrig, and J.~M.~{\'{A}}.
  Rodr{\'{\i}}guez.
\newblock Validating and describing linked data portals using {RDF} shape
  expressions.
\newblock In {\em Proceedings of the 1st Workshop on Linked Data Quality
  co-located with 10th International Conference on Semantic Systems,
  LDQ@SEMANTiCS 2014.}, volume 1215 of {\em {CEUR} Workshop Proceedings}.
  CEUR-WS.org, 2014.

\bibitem{Motik07}
B.~Motik, I.~Horrocks, and U.~Sattler.
\newblock {Adding Integrity Constraints to OWL}.
\newblock In C.~Golbreich, A.~Kalyanpur, and B.~Parsia, editors, {\em OWL:
  Experiences and Directions 2007 (OWLED 2007)}, Innsbruck, Austria, June 6--7
  2007.

\bibitem{Patel-Schneider2015}
P.~F. Patel-Schneider.
\newblock Using description logics for rdf constraint checking and closed-world
  recognition.
\newblock In {\em Proceedings of the Twenty-Ninth AAAI Conference on Artificial
  Intelligence}, AAAI'15, pages 247--253. AAAI Press, 2015.

\bibitem{Perez09}
J.~P{\'e}rez, M.~Arenas, and C.~Gutierrez.
\newblock Semantics and complexity of sparql.
\newblock {\em ACM Trans. Database Syst.}, 34(3):16:1--16:45, Sept. 2009.

\bibitem{EricRecordClasses15}
E.~Prud'hommeaux and J.~E. Labra~Gayo.
\newblock {RDF} ventures to boldly meet your most pedestrian needs.
\newblock {\em Bulletin of the American Society for Information Science and
  Technology}, 41(4):18--22, 2015.

\bibitem{EricSemantics2014}
E.~Prud'hommeaux, J.~E. Labra~Gayo, and H.~R. Solbrig.
\newblock Shape expressions: an {RDF} validation and transformation language.
\newblock In {\em Proceedings of the 10th International Conference on Semantic
  Systems, {SEMANTICS} 2014}, pages 32--40. {ACM}, 2014.

\bibitem{Organization14}
D.~Reynolds.
\newblock {The Organization Ontology}.
\newblock \url{http://www.w3.org/TR/vocab-org/}, 2014.

\bibitem{OSLCResourceShapes}
A.~G. Ryman, A.~L. Hors, and S.~Speicher.
\newblock {OSLC} resource shape: A language for defining constraints on linked
  data.
\newblock In C.~Bizer, T.~Heath, T.~Berners-Lee, M.~Hausenblas, and S.~Auer,
  editors, {\em Linked data on the Web}, volume 996 of {\em CEUR Workshop
  Proceedings}. CEUR-WS.org, 2013.

\bibitem{Simister13}
S.~Simister and D.~Brickley.
\newblock Simple application-specific constraints for rdf models.
\newblock In {\em RDF Validation Workshop. Practical Assurances for Quality RDF
  Data}, Cambridge, Ma, Boston, September 2013. W3c,
  \url{http://www.w3.org/2012/12/rdf-val}.

\bibitem{Slawek2015}
S.~Staworko, I.~Boneva, J.~E. Labra~Gayo, S.~Hym, E.~G. Prud'hommeaux, and
  H.~R. Solbrig.
\newblock {Complexity and Expressiveness of ShEx for {RDF}}.
\newblock In {\em 18th International Conference on Database Theory, {ICDT}
  2015}, volume~31 of {\em LIPIcs}, pages 195--211. Schloss Dagstuhl -
  Leibniz-Zentrum fuer Informatik, 2015.

\bibitem{Tiao10}
J.~Tao, E.~Sirin, J.~Bao, and D.~L. McGuinness.
\newblock Integrity constraints in {OWL}.
\newblock In {\em Proceedings of the 24th AAAI Conference on Artificial
  Intelligence (AAAI-10)}. AAAI, 2010.

\bibitem{Zaveri2016}
A.~Zaveri, A.~Rula, A.~Maurino, R.~Pietrobon, J.~Lehmann, and S.~Auer.
\newblock Quality assessment for linked data: A survey.
\newblock 7(1).

\end{thebibliography}


\end{document}